# On Backhauling of Relay Enhanced Networks in LTE-Advanced

Ömer Bulakci

Aalto University School of Electrical Engineering, Espoo, Finland, omer.bulakci@ieee.org

*Abstract*— Relaying is considered a promising cost-efficient solution in 3GPP LTE-Advanced for coverage extension and throughput enhancement. The compact physical characteristics and low power requirements of the relay nodes offer more flexible deployment options than traditional macro evolved Node Bs. This paper provides an overview of general relaying concepts and presents the relay deployment within the LTE-Advanced framework. Furthermore, the impact of relay backhauling on envisioned relaying gains is discussed and the methods to improve the performance of the backhauling are included.

## I. INTRODUCTION

LONG Term Evolution-Advanced (LTE-Advanced) is the candidate technology of the 3rd Generation Partnership Project (3GPP), which defines the framework for further advancement in LTE to fulfill the requirements of International Mobile Telecommunications Advanced (IMT-Advanced) specified by International Telecommunication Union-Radiocommunication (ITU-R). In accordance with these requirements, LTE-Advanced should support peak data rates of 1 Gbps on the downlink (DL) and 500 Mbps on the uplink (UL), bandwidth scalability up to 100 MHz, increased spectral efficiency up to 15 bps/Hz in UL and 30 bps/Hz in DL, along with improved cell edge capacity, as well as decreased user and control plane latencies relative to LTE Release 8 (Rel. 8) [1]. In order to meet these requirements, problems such as low signal-to-interference-plus-noise-ratio (SINR) at the cell edge and coverage holes due to shadowing and non-line-of-sight (NLOS) connections should be tackled.

The expected high data rate transmission with the future wireless communication networks necessitates upgrades for the current network paradigm. As the link performance of LTE Rel. 8 is already very close to the Shannon limit [2], new deployment topologies are taken into account. An option is to significantly increase the density of evolved Node Bs (eNBs). However, this implies high deployment costs and it is unlikely that the number of subscribers increases at the same rate, which turns out to be unappealing for network operators. A promising solution is deploying decode-and-forward relay nodes (RNs) near the cell edge. This type of RNs do not suffer from the limitations like loop back interference between transmit and receive antennas and noise/interference enhancement which are typical issues for amplify-and-forwards RNs [3]. Unless otherwise stated, decode and forward RNs are assumed in the rest of the paper.

RNs are relatively small nodes with low power consumption, which connect the core network with wireless backhaul through a donor eNB. This feature offers deployment flexibility and eliminates the high costs of a fixed backhaul link (eNB-to-relay link). It is expected that RNs improve cell edge capacity, lower operational expenditures (OPEX), reduce backhaul costs and enhance network topology [4][5].

Principally, a RN can almost be considered a wireless eNB, which include functionalities such as radio resource management, scheduling and hybrid-ARQ (HARQ) retransmissions [6]; however, the backhaul link is based on the LTE air interface rather than a fiber or microwave interface. The backhaul link can either use an additional frequency band (out-of-band RNs) or operate in the same spectrum as communication from/to mobile terminals (in-band RNs). The in-band RNs are universally deployable, since they do not require additional frequency licenses. Moreover, in [7] it is shown that in a coverage limited scenario, in-band RNs perform almost equally well compared to out-of-band RNs in terms of tackling both throughput and coverage gaps on the cell edge. Despite the advantages of in-band relaying, a proper backhaul support becomes more crucial. The constraints concerning backward compatibility to LTE, trade-offs between backhaul capacity and access capacity, and control channel designs should be taken into account.

In this paper, an overview of the basic relaying concept as well as the relay deployment within the LTE-advanced framework is provided. This is followed by the discussion on the impact of the backhauling on the RN deployments, where implementation issues along with backhaul optimization strategies are presented. Finally, a conclusion is given.

## II. RELAYING FUNCTIONALITY

### A. Relaying concept

The use of radio relaying for capacity enhancement and high data rate coverage extension has been discussed in academia for a long time [8][9]. The earlier studies on relaying were rather theoretical and focused on the network information theory aspect. In [10], Cover and El Gamal formulated capacity theorems for a simple relay channel. Moreover, multiple-input multiple-output (MIMO) techniques for relay networks are also considered and capacity bounds for relay MIMO channels are studied [11].

The relaying functionality can be realized either in a cooperative or multi-hop fashion. The cooperative use of relays creates virtual transmit diversity and exploits the spatial separation resulting in substantial increase in the available capacity [9]. Fig. 1 illustrates a simple scenario where a message is transmitted by a source node to the destination node both through a cooperative relay and directly. The destination node then combines the signals received from the relay and the source node and can exploit the diversity gain. On the other hand, in multi-hop relaying the source node communicates with the destination node either directly or via a relay, not both at the same time. In Fig. 2, the most typical benefits of the multi-hop relaying are shown.

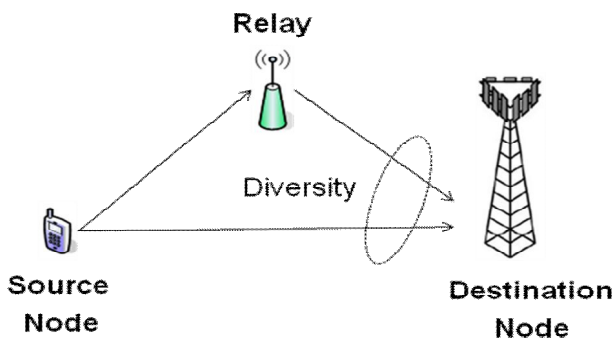

Figure 1. Example scenario for the cooperative relaying.

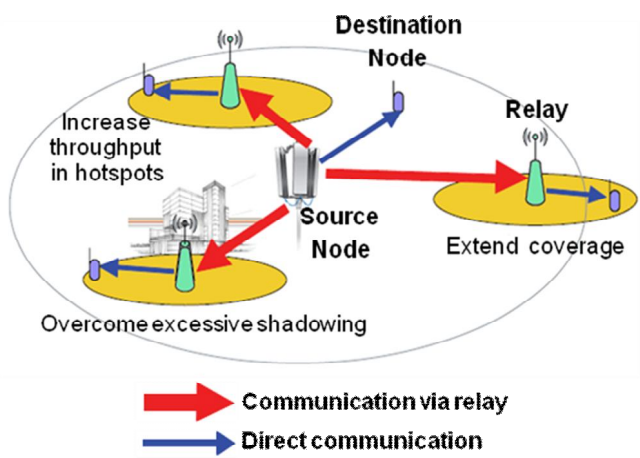

Figure 2. Benefits of the multi-hop relaying.

### B. Relay deployment in LTE-Advanced

Following the maturity of the digital wireless technologies and the drastic increase in the demand for high data rate coverage, relaying has found its way into the pre-standardization activities like IST-WINNER project and IEEE 802.16j standard which specifies relaying for the mobile WiMAX (802.16e) systems [12]. In addition, relaying has been investigated within the study item phase of LTE-Advanced as a technology to enhance coverage and capacity and to enable more flexible deployment options at low cost. Recently, the relay work item was accepted to specify in-band relays (type 1 relays) at least for the coverage-improvement scenario [13]. Therefore, the RNs are deployed at the cell edge to provide coverage. An example two-hop RN deployment is depicted in Fig. 3 where eNB stands for evolved Node B. Note that, since in-band RNs are considered, the backhaul link operates in the same carrier frequency as the relay-to-user link (access link).

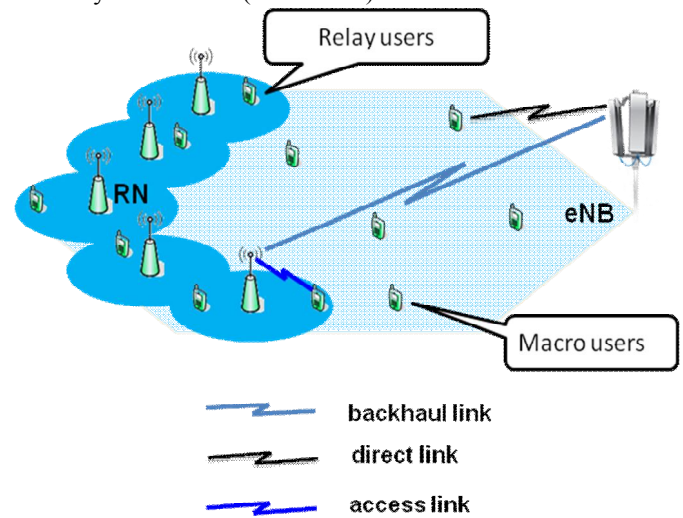

Figure 3. RN deployment at the cell edge. Relay users are served via RNs, whereas macro users are directly served by the donor eNB. Such a deployment offers coverage extension, where the cell edge users are connected to the RNs experiencing less path loss and benefiting higher resources.

A type 1 relay is described by the following [14]:

- A relay cell appears as a separate cell distinct from the donor cell to user equipments (UEs).

- The relay cells have their own physical cell IDs, i.e. a UE can synchronize to a RN directly during the cell search.

- In the context of single-cell operation, a UE is connected either to the donor eNB or a RN, but not both. In addition, the UE shall receive scheduling information and HARQ feedback from the RN and send its control channels to the RN.

- In the context of backward compatibility, a RN appears as an LTE Rel. 8 eNB to LTE Rel. 8 UEs so that all legacy LTE Rel. 8 UEs can be served by the RN. On the other hand, the RN should appear differently than the LTE Rel. 8 eNB to LTE-Advanced UEs.

## C. Backhauling aspects: From theory to practice

The impact of the backhaul link can be modeled via end-to-end (e2e) throughput. The optimal e2e expression of a relay UE (in-band or out-of-band) is given by the parallel formula[1]:

$$T_{e2e}^{Opt} = T_{access} // T_{backhaul} = \left(\frac{1}{T_{access}} + \frac{1}{T_{backhaul}}\right)^{-1}, (1)$$

where $T_{access}$ and $T_{backhaul}$ are the throughput levels on access link and backhaul link, respectively [7][15]. Note that, this is the two-hop specific expression. It can be derived that the e2e throughput converges to the throughput on the access link given that $T_{backhaul} >> T_{access}$. Therefore, the e2e throughput of an out-of-band relay UE converges to the access throughput and the backhaul TP has no effect, whereas the backhaul throughput has a key role for an in-band UE.

Beside the advantages of the in-band relaying, the backhaul support becomes crucial for a proper relaying operation within the specified LTE-Advanced framework as discussed in the previous section. First, simultaneous communication with the donor eNB and relay UEs is not desired in order to prevent interference between relay transmitter and receiver. A solution is reserving some subframes for the backhaul transmission, or in other words creating gaps in the relay-to-UE transmission. For instance, for downlink (DL) transmission during the reserved subframes, the RN is not able to transmit to the UEs, because it receives data from the eNB on the DL band. Such a gap structure can be realized either by blank subframes [16] or multicast-broadcast single-frequency network (MBSFN) subframes [17]. As a matter of fact, to support LTE Rel. 8 UEs the gap structure should also exist in LTE Rel. 8. Therefore, MBSFN mechanism which is already included in LTE Rel. 8 has been accepted to support the backhaul traffic [14]. During an MBSFN subframe the UE will not expect transmission except the reference and control signals that can be appended to the beginning of the subframe. These signals are expected in all subframes by the UEs to manage efficient synchronization, demodulation and mobility related measurements. Consequently, the remaining transmission gap can be used for the eNB-to-relay communication. An example relay-to-UE communication using the MBSFN subframe is illustrated in Fig. 4. Another issue which arises from MBSFN mechanism is that the RN is not able to receive the control and reference symbols. Hence, the most straightforward approach is to transmit the control channels for the backhaul link in subsequent symbols, i.e. during the transmission gap. Relay specific control channel transmission then implies additional overhead on the backhaul link.

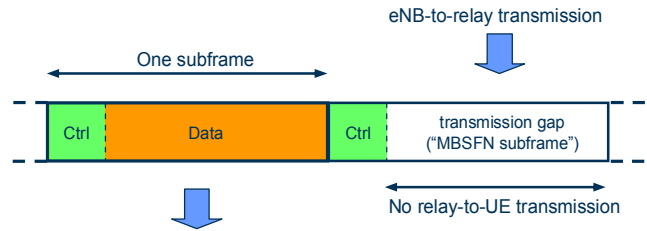

Figure 4. Illustration of relay-to-UE transmission using normal subframes (left) and eNB-to-relay transmission using configured MBSFN subframes (right) [14].

One drawback of the in-band relaying compared to the out-of-band relaying is the relatively decreased capacity because of the half-duplex feature. Although the relay users benefit from the higher resources and decreased path-loss due to the proximity to the RN, the access capacity strongly depends on the backhaul capacity. The backhaul link quality can be increased by several ways:

- Proper site planning techniques yield significant signal-to-interference-plus-noise-ratio (SINR) gains on the backhaul link along with a clear reduction in the shadowing standard deviation compared to random deployment. The aim of these techniques is to find an optimum RN location and site such that the backhaul link is not impacted by the shadow fading and hence, the SINR on the backhaul link is optimized [18]. An illustration of the basis of the relay site planning is shown in Fig. 5. In this illustration, the most favorable RN location is RN4, as the RN does not experience shadowing towards the donor eNB while the interfering eNB is shadowed.

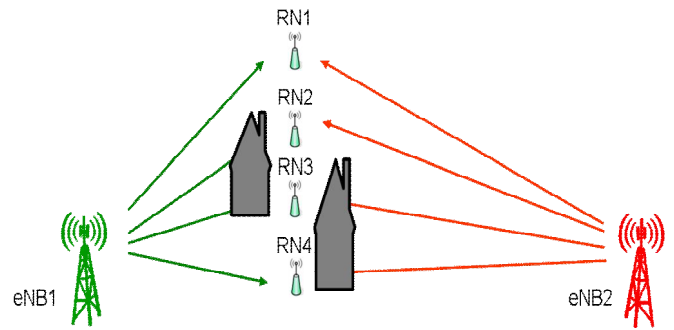

Figure 5. The impact of site planning on the backhaul link quality. The closest eNB on the left is typically the donor, and the other eNB is typically the interferer. The shadow fading is visualized by the houses.

- Using a set of directional antennas pointing toward the donor eNB leads significant increase in the link quality [19]. However, using additional directional antenna set increases the cost.

- For fixed RN locations, space-division multiple access (SDMA) can be utilized to improve the

---

[1] This expression is valid when the resource partitioning is optimized and equal amount of information is transferred over each link [15].

spectral efficiency of the backhaul link. In this scheme multiple beams are used for the backhaul transmission, where the direction of each beam is aligned according the location of the intended RN. Hence, multiple RNs can be served simultaneously within the cell. Although, the interference levels on the backhaul link are increased, it is shown that the multiplexing gain achieves significant improvement in access capacity due to increased capacity on the backhaul link [20]. Such a scenario is presented in Fig. 6.

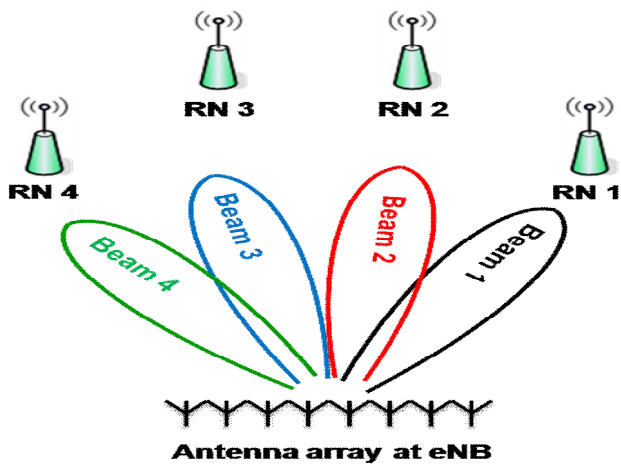

Figure 6. SDMA example. Multiple beams are directed to multiple RNs within the cell.

- In [19] it is shown that via a simple backhaul/access subframe partitioning, the gains of in-band relaying in sector and cell-edge throughputs are not significant. Therefore, advanced resource partitioning schemes can further improve the backhaul link.

In recent studies, it has been shown that relay enhanced LTE-Advanced networks offer considerable throughput gains compared to macro eNB-only scenarios. However, the backhaul link capacity seems to limit the achievable gains and thus the backhaul link becomes the bottleneck especially for the sector throughput [20][21]. On the other hand, for the cell edge throughput which is decisive for the cell coverage, the backhaul link has less impact given that the backhaul throughput is much higher than that of the cell edge throughput [15].

## III. CONCLUSION

Relaying is a promising enhancement to current radio technologies, which has been recently considered in the 3GPP LTE-Advanced study item and work item. In this paper, an overview of the relaying technologies as well as the relaying concept within the LTE-advanced framework is given. Possible gains of relaying are discussed and backhauling is found to be critical to achieve the envisioned gains of the relaying. Moreover, the methods to improve the backhaul link are presented.